\documentclass[review]{elsarticle}

\usepackage{hyperref}
\usepackage{lineno,hyperref}
\usepackage{booktabs}
\usepackage{graphicx,calc}
\newlength\myheight
\newlength\mydepth
\settototalheight\myheight{Xygp}
\usepackage{comment}
\usepackage{wrapfig}
\usepackage{subcaption}
\usepackage[font=small]{caption, subcaption}
\usepackage{todonotes}

\usepackage[section]{placeins}
\usepackage[misc,geometry]{ifsym}

\journal{} %Journal of \LaTeX\ Templates}

\begin{document}

\begin{frontmatter}

\title{Defining and Detecting Toxicity on Social Media: Context and Knowledge are Key}

\author{Amit Sheth\textsuperscript{1}}%\fnref{myfootnote}}

\author{Valerie L. Shalin\textsuperscript{2,1}}

\author{Ugur Kursuncu\textsuperscript{1,3}} %\corref{mycorrespondingauthor}}

\address{\textsuperscript{1} AI Institute, University of South Carolina}
\address{\textsuperscript{2} Department of Psychology, Wright State University}

\address{\textsuperscript{3} J. Mack Robinson College of Business, Georgia State University}

\begin{abstract}
Online platforms have become an increasingly prominent means of communication. Despite the obvious benefits to the expanded distribution of content, the last decade has resulted in disturbing toxic communication, such as cyberbullying and harassment. Nevertheless, detecting online toxicity is challenging due to its multi-dimensional, context sensitive nature. As exposure to online toxicity can have serious social consequences, reliable models and algorithms are required for detecting and analyzing such communication across the vast and growing space of social media. In this paper, we draw on psychological and social theory to define toxicity. Then, we provide an approach that identifies multiple dimensions of toxicity and incorporates explicit knowledge in a statistical learning algorithm to resolve ambiguity across such dimensions. 
\end{abstract}


\end{frontmatter}

%\linenumbers

\section{Introduction}
\label{sec:intro}
\vspace{-1mm}
Online social media platforms are arguably among the most culturally significant technological innovations of the 21st century. The numerous benefits include the wide distribution of content crossing geographic boundaries, and enabling interaction and exchanges that are nearly free of physical constraints except for infrastructure. Communities have emerged around every conceivable special interest from science to travel, from politics to child-rearing. The easy spread of data, information, and knowledge was expected to foster informed decision-making, cultural exchanges, and the coordination of activities online and in the physical world. Unfortunately, social media has also significantly enhanced the reach and scale of harmful content including disinformation, conspiracies, extremism, harassment, violence, and other forms of socially toxic material. While social media platforms attempt to counter such harmful content, their efforts are largely ineffective and as such themselves have the potential for unintended adverse impact. The effectiveness of moderation is potentially biased by the platforms’’ economic interest or political and regulatory considerations.  Or failure may simply be due to the lack of effective tools and sufficient investment.  Irrespective of the reason, human content moderation has resulted in relatively unsatisfactory outcomes \cite{young2021much}. Although the political and public health climate of 2020 encouraged society to adopt technological and specifically AI-based solutions, success was also limited. A prominent reason is the lack of understanding of the challenging nature of toxicity, which fundamentally requires context outside of the explicit content. 
The detection of toxicity demands an interdisciplinary perspective with empirical approaches. Consistent with our people-content-network framework for the characterization of social media exchange \cite{nagarajan2010qualitative,sheth2010understanding,purohit2011understanding,kursuncu2018s}, we assert the more general role of context, and in particular cultural context, in the interpretation of content. This paper has  three goals:

\begin{itemize}
    \item identify the  psychological and social dimensions of the problem
    \item identify  the limitations of contemporary computational approaches, and
    \item outline an advanced technical approach founded on knowledge-driven context-based analysis.
\end{itemize}

\section{A PsychoSocial problem Meets Computation}
\label{sec:RW}
\begin{wrapfigure}{r}{6cm}
%\vspace{-1em}
  \begin{center}
    \includegraphics[width=0.32\textwidth, trim=2.5cm 2.5cm 3cm 1cm]{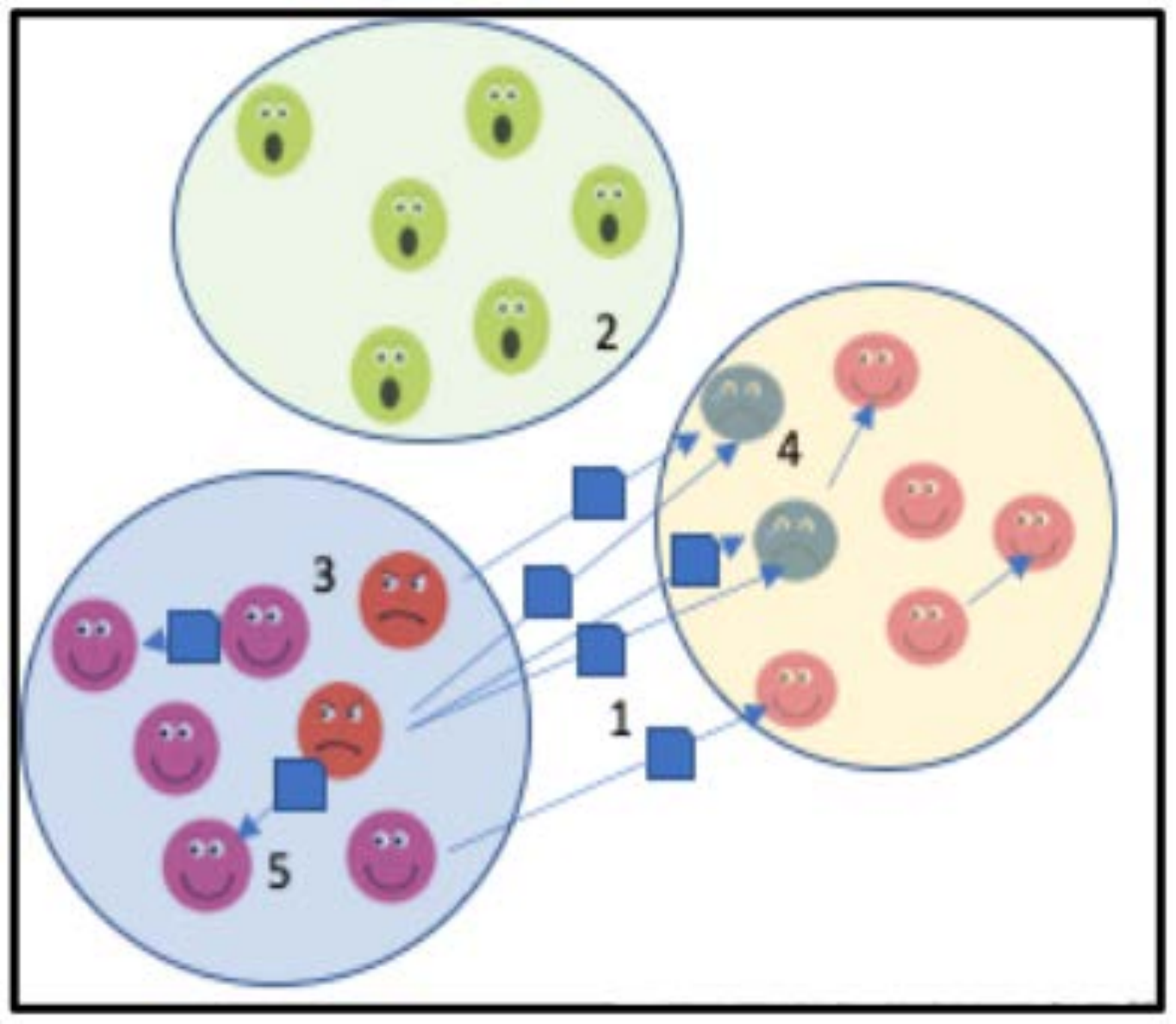}
  \end{center}
  \vspace{1em}
  \caption{\footnotesize Conventional toxicity analysis examines the content exchanged between individuals in a community (1). Often external observers impose their own culturally biased decision rules (e.g., gender-based violence) (2). Detecting toxic sources (3) expands analysis, but still fails to acknowledge the reaction of the target (4), which is likely tempered by common group membership (5).}
  \label{fig:figure-1}
%\vspace{2em}
\end{wrapfigure}

\vspace{-2mm}
Our view of toxic content extends beyond the current classifications that focus on “threats, obscenity, insults, and identity-based hate”\footnote{https://www.kaggle.com/c/jigsaw-toxic-comment-classification-challenge/overview}. We also include harassment and socially disruptive persuasion, such as misinformation, radicalization and gender-based violence \cite{purohit2015gender,kursuncu2019modeling,rezvan2020analyzing,wijesiriwardene2020alone,wang2014cursing}.  While the cultural foundations of toxicity are readily apparent in misinformation and radicalization, we contend that culture provides essential context to the determination of any toxic content. Figure \ref{fig:figure-1} guides this section, starting with conventional content analysis and expanding the psychological, social, and cultural scope of the required analysis.

\subsection{Content Analysis}
By far the most common approach to toxicity detection focuses on the content of exchanges. Offensive keywords, often so-called “coarse language” are easy to tabulate in a lexicon.  More sophisticated analyses employ lexicons specific to intelligence, appearance, race,  sexual preference, etc. \cite{rezvan2020analyzing}. Keyword-based content analysis encounters a number of challenges. An evolving culture conveys an insulting connotation to otherwise apparently banal language, e.g., basic, cancel, Karen, shade, snowflake, and thirsty. This not only requires constant maintenance of the lexicon, but context to disambiguate the slang usage from general usage \cite{sheth2016semantic}.

A second problem is that content analysis based on isolated lexical items does not necessarily confer toxicity \cite{wang2014cursing,wijesiriwardene2020alone,kursuncu2021bad}. For example, North American teenagers readily employ language among themselves that adults would consider offensive. More worrisome, the word “jihad” may be readily interpreted as a radical content by a Westerner, but a more culturally sensitive analysis reveals that this term also appears in benign religious text. The scope of toxic topics, general knowledge, and cultural foundations required for interpretation is virtually unbounded.  “Dressing like your grandmother” directed to a teen is laden with cultural deprecation founded on both ageism and consumerism but contains no single offensive word in isolation. Irony, humor,  and teasing between friends precludes simple sentiment analysis.

Toxic content is also multimodal, often exploiting images and videos that are not well monitored or understandable using contemporary technology.  Facial recognition algorithms fail miserably for dark-skinned females. More generally, the benchmark image databases heavily favor western culture, at the expense of eastern cultures (China is represented by only 1\% of images in the Imagenet\footnote{https://devopedia.org/imagenet}). Apart from these obvious multi-modal processing challenges, text and image content must be aligned in a common framework at the appropriate level of abstraction.  Finally, we cannot assume that text provides useful image processing guidance.  A cheery “Have a Nice Day” text can easily pair with an embarrassing photo of the recipient. 

\subsection{Culturally Bound Decision Criteria}
Even a simple content analysis requires a decision criterion.  Framing toxicity as a  standard signal detection problem acknowledges the potential for two overlapping distributions of potentially toxic content instances, one over relatively low toxicity values and another over higher values. The decision criterion is vulnerable to cultural considerations.    All-too-common annotator disagreement is resolved by the vote of a small sample of annotators, while the foundations of disagreement remain unstudied.  Annotators may hold unconscious stereotypes, for example associating religion with radicalization.  Personal experience and cultural differences create variable interpretations of label semantics; disagreement over interpretation is the source of great concern in moderating internet content. 

We have already noted the preponderance of coarse language in teenagers \cite{jay1992cursing,jay2009utility,mehl2003sounds}. The population of truly non-toxic instances is much larger than the population of toxic instances \cite{wijesiriwardene2020alone,waseem2018bridging,golbeck2017large}.  Such class imbalance does not work well with contemporary machine learning algorithms.  As we lower the toxicity decision rule we admit more false positives, potentially resulting in orders of magnitude more false positives than true positives.  This creates both adverse impacts to the falsely accused along with a practical problem of follow up \cite{kursuncu2019modeling}. 

Finally, corpus assembly itself conveys cultural bias.   The classification algorithm for one population does not generalize to different populations, creating a validity problem.  As implied above, the relevant data set for detecting toxic exchange among adolescents differs from the data set for defining such exchanges among adults \cite{wijesiriwardene2020alone} due to different cultural practices.

\subsection{Identifying the toxic source}
The source’s intent to hurt or harm is the defining feature of bullying. Harm may employ the disclosure of sensitive facts, denigrate, be grossly offensive, be indecent or obscene, be threatening, make false allegations, deceive, spam, spread misinformation, mimic interest, clone profile or personal invade space. The motivation for detecting the toxic source is mitigation, but the false alarm risk is real.  Moreover, one instance is unlikely to constitute sufficient evidence.  Evaluation of the potentially toxic source requires a corpus of the candidate’s content, raises a challenge of corpus scope, and introduces the need for aggregation of evidence.

\subsection{Identifying the Target} 
Experienced harm is distinct from intent to harm, from the recipient’s perspective resulting in discrimination, deception, fraud, disinformation, loss of money, offense, loss of reputation, manipulation, embarrassment, distraction, loss of time. Moreover, harassment, by definition, refers to the special case of bullying with respect to a protected class \cite{einarsen2010bullying}.  The nefarious source takes advantage of the specific features of target vulnerability such as age, occupation, and public stature.  Contemporary victims of bullying include Parkland High School Students \footnote{https://www.huffpost.com/entry/parkland-students-combat-cyberbullying\_n\_5aeeee86e4b033e5c3f03126}. The motivation for detecting the target is protection. Both bullying and harassment are associated with cases of adolescent suicide \cite{hinduja2010bullying,roberts1996strategies,cooper2012examining}. Age most certainly matters in the assessment of target experience; adolescent brains are still developing an ability to process social feedback making adolescents particularly vulnerable to both negative feedback \cite{crone2018media},  and radicalization efforts \cite{pedersen2018risk}. 

\subsection{Participants’ Relationship and Group Membership}
Friendship, power differentials, and social network membership provide essential context.    Insults are common among adolescent friends. Participants with the same racial background readily exchange otherwise offensive racial epithets. Social network structure has at least two consequences founded on the distinction between in-group and out-group membership and the target’s position within these groups.   First, multiple negative messages from different participants in the in-group targeted to an out-group recipient are as potentially toxic as the same number of messages from a single source. Social network membership is therefore an important feature in the detection of toxicity.  Second, the promise of group membership and the threat of exclusion is a known factor in the radicalization effort \cite{ozer2020group}, of particular appeal to adolescent recruits. Thus, network-driven dynamics play an important role in the detection of toxicity, especially when toxicity is amplified by networked efforts such as hate campaigns and organized misinformation spread.

Because surrounding benign conversation mitigates the single potentially toxic comment, exchange history informs the determination of toxicity. Hence, an exchange history surrounding the potentially toxic comment must be present in the corpus to enable evaluation. Crawling on sender and recipient identifiers is too limited.  Victims are often targeted with mention tags in an exchange between a sender and what might be charitably called bystanders.  These concerns illustrate that the scope of a purportedly toxic item influences the annotation task. A single episode may look quite different in the context of other exchanges, suggesting that the potentially benign or toxic instance should be annotated with respect to its broader historical and network context. This argues for new requirements for systematically defining and scoping the annotation task. Expanded context also raises the problem of conflicting indicators, the assessment of stale content, and the need for confidence estimates.

\section{Technical Challenges to Automated Detection of Toxic Language}
\label{sec:dataset}
\vspace{-2mm}
As described above, the toxicity detection problem is not a purely computer science or AI problem. To identify toxicity, it is necessary to understand the broader context beyond the situation and domain-specific content analysis, with reference to applicable human values, social norms, and culture, at the individual, group, and community levels \cite{purohit2020knowledge}. Toxicity detection is an interdisciplinary problem founded on theory, empirical models, and knowledge to guide classification \cite{henry2009school,wijesiriwardene2020alone,kursuncu2021bad,kursuncu2020cyber,kursuncu2021cyber}. In contrast, conventional approaches for the identification of toxic exchange have been treated as a  content processing problem \cite{noever2018machine,pavlopoulos2020toxicity}. Researchers often rely on the post-level for building datasets and designing algorithms to detect toxicity between two individuals relying on the explicit language of insult \cite{wijesiriwardene2020alone}. The state-of-the-art algorithms used to model toxic content are mostly autoregressive models (e.g., BERT, GPT-2,3),  designed to predict the next token given previous tokens from the dataset as input. As these models have been trained using data collected from the web, corpus bias and incidentally confounded features result in models that can cause intentional or unintentional harm\footnote{ https://thenextweb.com/neural/2021/01/19/gpt-3-has-consistent-and-creative-anti-muslim-bias-study-finds/} \cite{mcguffie2020radicalization,olteanu2019social,kursuncu2021bad}. For example, recent studies \cite{gehman2020realtoxicityprompts,groenwold2020dats,wallace2019universal} suggested that these state-of-the-art algorithms are prone to generating racist or sexist schemes. While these models can be retrained using transfer learning by fine-tuning model parameters, significant harmful bias will still carry over.  Such models can be dangerous in highly consequential areas, such as online toxicity as well as health \cite{zhang2020hurtful,garg2018word,chen2019can}. For instance, Google’s Perspective API\footnote{https://www.perspectiveapi.com/} designed for toxicity detection received criticism for biased scoring of content based on gender, sexual orientation, religion, or disability. Further, this model almost always assigned a high toxicity score if the content included insults or profanity, regardless of the intent or tone of the author \cite{hanu2021aitoxic}. Hence, policymakers and practitioners justifiably assert serious usability and safety concerns that constrain the adoption of poorly understood technologies \cite{topol2019high}. 

Below, we discuss the technical consequences of our expanded approach to toxicity detection in three subsections: the need for empirical models, the need for a curated corpus, and the need for external knowledge. 

\subsection{Need for Empirical Models}

Computational modeling of human behavior requires domain expertise to inform the classes and subclasses of toxicity. On the other hand, such domain expertise is scarce; hence, we require a conceptual model in a structured or semi-structured format that is readable by machines.  The presently available models are impoverished.  Below we consider the use cases for cursing, extremism, and harassment to demonstrate the need for better empirical models to guide content analysis. 

\paragraph{Cursing} The intention of the parties of a conversation along with social context determines the meaning of their language. In \cite{wang2014cursing}, we studied the communications on Twitter concerning the use of cursing and its relations with intention and emotions. While we found around 8\% of conversations contain profanity and curse words, the intention of users may not necessarily be toxic. We explored the role of emotions in identifying intention, as cursing may be associated with positive emotions as well as negative emotions, and these emotions may indicate the real intention. We identified three variables outside of the text that determines \emph{when, where} and \emph{how} cursing occurs. For example, we found that people curse more when they wake up, in relaxed virtual environments. Clearly, such features of social context should moderate the interpretation of cursing as a toxic indicator. 

\paragraph{Extremism} In \cite{kursuncu2019modeling,kursuncu2018modeling,arpinar2016social} we started with the political science notion that radicalization is a process employed by extremist groups, with systematic changes in persuasive content over time. For the particular use case of Islamist extremism, appropriate domain expertise is critical to distinguish the true extremist from non-extremist communication. Guided by an empirical model developed by a political scientist, we examined three dimensions to model this content: religion, ideology, and hate. Ambiguity is a significant challenge. For instance, the meaning of the keyword “jihad” in religion is referred to as a self-spiritual struggle, while it indicates intent to harm other individuals in the Islamist extremist ideology. As the same term has two different meanings for extremist and non-extremist content, it needs to be represented differently in a computational model for resolving such ambiguity. Hence, a multi-dimensional and contextually sensitive model of this content incorporating knowledge (in this example, religious knowledge) allows us to address ambiguity, reducing false alarms and mitigating unfairness. Such a socially responsible model with improved fairness would mitigate the adverse impacts on nearly 2 billion Muslims.

\paragraph{Harassment} Many early researchers defined harassment as a binary classification - a social media post (e.g., a tweet) is either harassing or not \cite{yin2009detection,kennedy2017technology,bastidas2016harassment,bugueno2019harassment}. As the context is crucial in capturing harassment, content classification will change based on the linguistic meaning, interpretation, and distribution. In \cite{rezvan2020analyzing} we expanded the  dimensions of harassment to include; (i) sexual, (ii) racial, (iii) appearance-related, (iv) intellectual, and (v) political content, and created a type-aware lexicon and annotated dataset \cite{rezvan2018quality}. Then we employed a multi-class classification algorithm based on these five dimensions. While coarse lexical items signal some of these, ambiguous common language (fat, dumb) and idioms are also relevant. A multi-class approach is required because perpetrators can exploit more than one subclass in targeting a victim.  Critically, in the absence of a multi-class model, the victim’s experience of harassment over time will not surface.

\subsection{Need for Curated Corpora}
The analysis is only as good as the corpus. As previously noted, researchers often resort to the post-level for building datasets and designing algorithms to detect toxicity between two individuals, focusing on recognizing the explicit language of insult \cite{wijesiriwardene2020alone}. Keyword-based crawls in corpus assembly create misleading corpora, rife with ambiguity, e.g., a playful exchange between good friends with sarcastic content could be falsely flagged as harassment, or a religious reference to “jihad” could falsely flag a pious worshipper. We next consider the corpus assembly problem for extremism and harassment. 

\paragraph{Extremism} For our extremism project, we relied upon a curated corpus \cite{fernandez2018understanding} consisting of 538 verified extremist users, established by Twitter and the Lucky Troll club \cite{ferrara2016predicting}.  We balanced this with a corpus of 538 non-extremist users from an annotated Muslim religious dataset \cite{chen2014us}. We make two points with this example. First, the set of positive cases reflected professional judgment. Second, the applicability of the resulting classification model depends on the quality of the distractor corpus. Here we were particularly concerned with adverse impact and therefore employed a distractor corpus that posed a significant false alarm opportunity. Nevertheless, this balanced corpus does not reflect the class imbalance in the uncurated data. Even very high precision results can produce a large number of false alarms in the natural,  unbalanced corpus \cite{kursuncu2019modeling}.  

\paragraph{Harassment} We curated our own corpus for our high school harassment that addresses a number of corpus considerations \cite{wijesiriwardene2020alone}, under an IRB-approved protocol requiring privacy protections through anonymization. This corpus, called ALONE, comprises of posts exchanged in interactions between pairs of participants to capture the appropriate contextual cues. Hence, each sample is an interaction that is an aggregate of posts between users along with other metadata. First, because the culture of the U.S. high school population is quite different from the general U.S. culture at large, we assured the identity of the participants. Starting with a seed set of known high school student names published in the newspaper as scholarship winners, we searched Twitter for unique matches to users with appropriate location indicators in their metadata.  To grow the set, we searched on their Twitter contacts and then pruned the resulting list of candidates by requiring contacts with other members of the candidate list.  Second, we make no assumptions regarding the nature of toxic content in assembling this corpus.  Third, as we were concerned with capturing the full context for the individual post, we retrieved the history of exchanges between members and the multi-modal content of these exchanges including emoji and images which may also contain toxic content \cite{wijesiriwardene2020alone}. The diversity of modality enriches the interactions between humans and computers. Specifically, users create the context for their conversations using these modalities. As a result of our corpus assembly process, we can recover network structure \cite{bhatt2019knowledge} suitable for insider-outsider analysis.  Finally, with the caveat of access restricted to public accounts, our corpus approximates a realistic class balance of benign and toxic content.

\subsection{Need for Computationally Accessible External Knowledge}

We advocate the use of relevant external knowledge in a variety of forms, including text sources and computationally accessible knowledge graphs (KGs). This assures attention to the different dimensions to account for subtle nuances in the semantics of toxic behavior. External knowledge constitutes a source of “ground truth” for evaluating message content.  As we argue that toxic behavior is multi-dimensional, leading to ambiguity and false alarms, we employ a multi-level and multi-dimensional approach that helps capture differences between various cultural and social senses of toxicity to resolve ambiguity.  Our previous Person, Content, and Network (PCN) distinction  \cite{purohit2011understanding,sheth2014twitris,kursuncu2018s,kursuncu2019predictive} functions at a higher (superficial) level, whereas the contextual dimensions of content (e.g., religion, ideology and violence) functions at the lower level \cite{kursuncu2019modeling}, capturing the deep semantics of toxicity. Further, when incorporated into a classification algorithm, external knowledge enables opportunities to provide an explanation generally missing from contemporary deep learning approaches \cite{wang2019explainable,lecue2020role,gaur2021semantics,bansal2021combining}. 

\begin{figure}
%\vspace{-4em}
    \centering
    \includegraphics[width=0.95\linewidth]{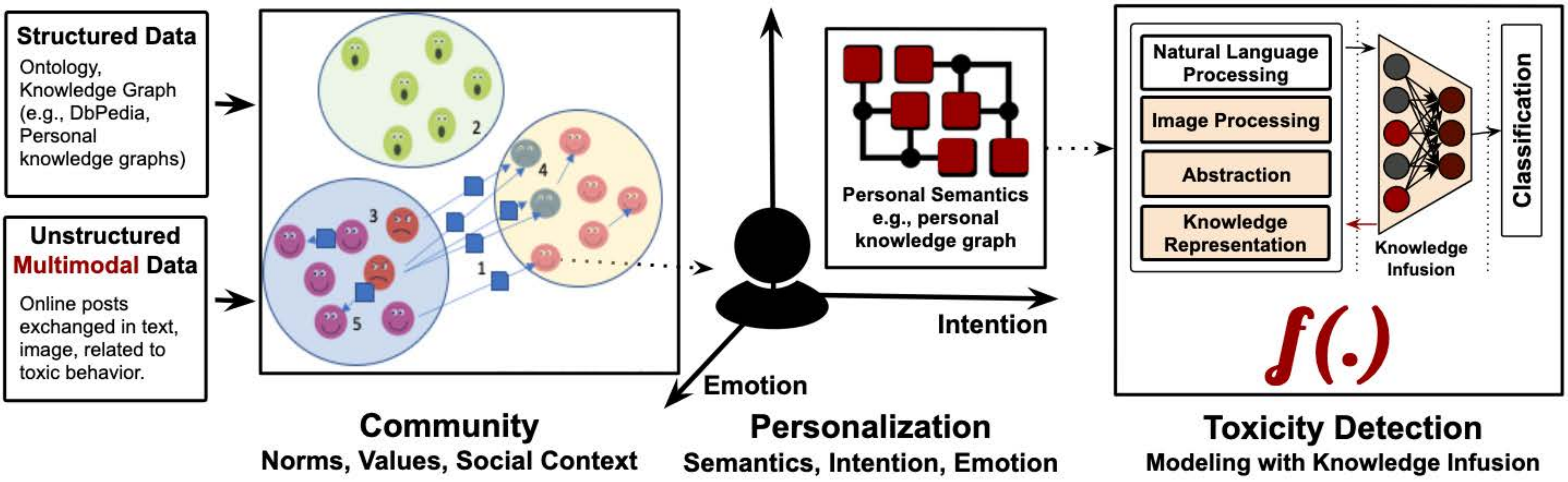}
\caption{\footnotesize Individuals are surrounded by the sources of data and knowledge required for computational analysis. Personalizing the analysis by incorporating personal semantics, intention, and emotion will help distinguish toxic behaviors from non-toxic. Further, infusing external knowledge will resolve ambiguity by better contextualizing multimodal data and providing a source of explanation.}
\label{fig:figure-2}
\vspace{-1.5em}
\end{figure}

Purohit et al. \cite{purohit2020knowledge}, provides a complementary framework for broader context to guide interpretation and evaluation. They identified three major dimensions of knowledge necessary to design humanity-inspired AI systems: personalization, social context, and intention. Here, we expand \cite{purohit2020knowledge} to scope the relevant knowledge that they described in three dimensions: values, norms, and the domain. Each dimension is pegged by individual specificity and collective generality, and the perspective required to interpret the behavior of an individual is represented by the combination of all three dimensions. While \cite{purohit2020knowledge} considered other actors as part of the environment, here, we consider them more explicitly (see Figure \ref{fig:figure-2}). They constitute a community, with norms and values. The concept of \emph{personal semantics} for the target of toxicity covers much of what \cite{purohit2020knowledge} intended in their analysis. Personal semantics includes knowledge about the targets’ language of insult, verbal abuse, and offensive language, involving sensitive topics specific to the individual and their social network. From the sources’ perspective, we require knowledge corresponding to their \emph{intention}, particularly associated with indicators of power, truth, and trust \cite{mayer1995integrative}. Finally, the \emph{history of interaction} such as duration and toxicity frequency between source and target requires knowledge about the structure of nominal conversation such as indicators of topic change and common ground that determine familiarity \cite{hutchby2017conversation}. The target’s \emph{emotional response} corresponds to the toxicity-specific emotion evoked in a recipient after reading messages, informed by conversation history and network membership. Toxicity detection requires a more sophisticated classification scheme beyond binary toxicity, referring to knowledge related to  the causes of experienced harm, embarrassment, loss of reputation, etc. as well as possible clinically relevant consequences such as depression and suicide \cite{kursuncu2021bad}.  While these sources of knowledge are typically not made explicit in toxicity analysis, the failure to make them explicit or acknowledge features corresponding to these contributes to disagreement among annotators and ultimately poor, and biased classification.

\section{A Knowledge-enhanced Socio-technical Approach to Toxicity Detection}
\label{sec:disc-conc}
\vspace{-2mm}
Toxicity detection takes the form of two problems: detection of the toxic source(s) and identification of the vulnerable victim. For both problems, we require more sophisticated Natural Language Processing (NLP) and Machine Learning (ML) methods to detect and use the features indicative of toxicity. Because the meaning of the content is personalized based on the belief system of the source and target, the semantic meaning needs to be computationally represented separately. Such personalized belief systems are critical for understanding how toxic behavior is interpreted differently by different individuals \cite{friedkin2016network}. These inter-related concepts and beliefs also evolve over time upon exposure to new information \cite{uso2016belief}. The question here is how one can computationally model the evolution of such complex social exchange. 
We advocate a Knowledge-infused Learning (K-iL) framework \cite{kursuncu2018modeling,kursuncu2019knowledge,sheth2019shades} where the model learns to recognize patterns of different meanings of toxic concepts from different perspectives to reduce ambiguity. However, the knowledge sources are not necessarily at the same level of granularity and abstraction. Accordingly, we categorized knowledge infusion \cite{kursuncu2018modeling,kursuncu2019knowledge,kursuncu2019modeling} as shallow, semi-deep, and deep infusion to resolve the impedance mismatch due to different representational forms and abstractions \cite{sheth2019shades}. Infusing knowledge is particularly important for overcoming the inescapable limitations and biases of data-driven processing \cite{sheth2021duality}. 

We propose a framework that will account for Purohit et al.’s dimensions \cite{purohit2011understanding} to generate richer representations including personal semantics, intention, emotion, history of interaction, and social context. This collection of information will require dynamic hybrid models for different modalities of data and knowledge representation. As behavioral models are dynamic and evolve, this framework should also allow for change. Further, validation of such an approach is also challenging and likely requires some form of experimentally controlled data collection to support supervised learning. The framework must address multiple levels of analysis, such as content, individual, and community, ensuring that the individual level details are changing as a function of interacting with their network. Communities form around various topics of interest through network interactions, where the shared content displays an intent attached with emotions. As learning concepts and grasping causal relations go beyond the data available, conceptual and probabilistic models can perform inference over hierarchies of structured representations \cite{tenenbaum2011grow}. 

Among the Purohit et al. dimensions \cite{purohit2011understanding}, personal semantics, interactions, and social context can be represented using both conceptual (e.g., knowledge graphs) and probabilistic models (e.g., language, image). External knowledge can be represented in structured (e.g., knowledge graph) and semi-structured forms (e.g., JSON) to inform computation. While knowledge can be acquired from data through various methods, dependence on data significantly limits the search space and extraction of the complete knowledge that is required to represent the complex nature of toxicity \cite{valiant2000robust}. Explicit structural relations in a knowledge graph constitute context and capture the intrinsic characteristics of this problem, which can be incorporated into a statistical learning algorithm (e.g., neural networks) to enhance the latent contextual space. This incorporation will adjust emphasis on sparse-but-essential and irrelevant-but-frequent terms and concepts, boosting recall without reducing precision \cite{kursuncu2019modeling,kursuncu2018modeling,kursuncu2019knowledge}. While probabilistic models (e.g., BERT, GPT-3, ResNet, Inception) have advanced in recent years, generating knowledge representations from knowledge graphs or similar structured forms of knowledge remains an open area for advancement. However,  a knowledge graph can be represented as embedding vectors including structural information of the graph, such as relationships. Existing methods, such as TRANS-E \cite{bordes2013translating}, TRANS-H \cite{wang2014knowledge}, and HOLE \cite{nickel2016holographic}, can  generate embeddings from a knowledge graph. The generated knowledge representation can then be infused within a probabilistic model.

In a learning architecture, represented knowledge can be infused through an attention mechanism and knowledge-based constraints or dependency relations between words in a sentence \cite{sheth2019shades}. Deep infusion of knowledge is still an open area of research, as we described in \cite{kursuncu2019knowledge,kursuncu2018modeling}. Deep infusion of knowledge combines the representation of structural knowledge graph content with a latent representation of data, quantifying the information loss and identifying the level of abstraction. The infusion of knowledge can take place after each epoch optimizing the loss function. In this architecture, for deep infusion, related functions add an additional layer that takes the latent vectors of the previous layers, and the knowledge embedding, merging them to output a knowledge infused representation. In this framework, as we utilize multiple dimensions to represent toxic behavior, an appropriate infusion of knowledge will form connections within the data resulting in better-contextualized representation. As our prior work suggests, infusion of knowledge mitigates unfair outcomes by reducing false positives that would lead to adverse social outcomes. \cite{kursuncu2019modeling,kursuncu2019knowledge,kursuncu2021bad}.

\section{Conclusion}
\label{sec:disc-conc}
\vspace{-2mm}
In this paper, we identified the multiple influences on the detection of toxic exchange beyond conventional content analysis. Our goal was to provide a framework that identifies and utilizes the multiple dimensions of toxicity and incorporates explicit knowledge in a statistical learning algorithm to resolve ambiguity. For toxicity detection, we provided a framework founded on behavioral and social theory. Specifically, we highlighted the significance of multi-level analysis of data, namely, content, individual, and community, and the numerous features necessary to determine toxicity. Knowledge representation and its infusion in a learning algorithm is an emergent solution for toxicity detection and related sets of similar problems.

\section*{Acknowledgement}
\vspace{-3mm}
This work is funded in part by National Science Foundation Award 1761931 Spokes: MEDIUM: MIDWEST: Collaborative: Community-Driven Data Engineering for Substance Abuse Prevention in the Rural Midwest. Any opinions, findings, and conclusions or recommendations expressed in this material are those of the author(s) and do not necessarily reflect the views of the National Science Foundation.

%\section*{References}

\bibliography{references}

\end{document}